\documentclass{elsart}

\usepackage{graphicx,amssymb}
\journal{Solid State Communications}

\newcommand{\rfig}[1]{Fig.~\ref{#1}}

\begin{document}

\begin{frontmatter}


\title{Epitaxial graphene}
\author[GT]{Walt A. de Heer\corauthref{cor1}}
\ead{deheer@electra.physics.gatech.edu}
\corauth[cor1]{Corresponding author.}
\author[GT,CNRS]{Claire Berger}
\author[GT]{Xiaosong Wu}
\author[GT]{Phillip N. First}
\author[GT]{Edward H. Conrad}
\author[GT]{Xuebin Li}
\author[GT]{Tianbo Li}
\author[GT]{Michael Sprinkle}
\author[GT]{Joanna Hass}
\author[CNRS]{Marcin L. Sadowski}
\author[CNRS]{Marek Potemski}
\author[CNRS]{Gerard Martinez}
\address[GT]{School of Physics, Georgia Institute of Technology, Atlanta, GA 30332, USA}
\address[CNRS]{Grenoble High Magnetic Field Laboratory, CNRS, Grenoble, France}

\title{}

\author{}

\address{}

\begin{abstract}
Graphene multilayers are grown epitaxially on single crystal silicon carbide. This system is composed of several graphene layers of which the first layer is electron doped due to the built-in electric field and the other layers are essentially undoped. Unlike graphite the charge carriers show Dirac particle properties (i.e. an anomalous Berry's phase, weak anti-localization and square root field dependence of the Landau level energies). Epitaxial graphene shows quasi-ballistic transport and long coherence lengths; properties which may persists above cryogenic temperatures. Paradoxically, in contrast to exfoliated graphene, the quantum Hall effect is not observed in high mobility  epitaxial graphene. It appears that the effect is suppressed due to absence of localized states in the bulk of the material.Epitaxial graphene can be patterned using standard lithography methods and characterized using a wide array of techniques.  These favorable features indicate that interconnected room temperature ballistic devices may be feasible for low dissipation high-speed nanoelectronics.
\end{abstract}

\end{frontmatter}



\section{Introduction}
Carbon nanotubes are prototypical of quasi-one dimensional graphene nanostructures. The approximate electronic structure of a carbon nanotube with diameter $D$ is understood starting from the graphene dispersion relation, i.e. the Dirac cone $E=\hbar v_0|k|$, and quantizing the angular momentum about the axis so that $E_n=\hbar v_0\sqrt{k_z^2+k_n^2}$, where $k_n=(n+a)\pi/R$, where $a$ is 0 or 1/2 depending on whether the nanotube is metallic or semiconducting. A metallic nanotube has two dispersionless bands that cross the Fermi level while a semiconducting nanotube has a band-gap $E_g=\gamma_0 a_0/R\sim0.4$eV$\cdot$nm$/R$. This property, that graphene nanostructures can be metallic or semiconducting depending on their shape carries over to nanopatterned graphene ribbons as shown below.

High purity multiwalled carbon nanotubes (as well as single walled nanotubes) were found to be room-temperature ballistic conductors \cite{Frank1998}. This property requires (at least) that electrons traverse the length of the nanotube without scattering. This discovery coincided with predictions of the effect by Ando \cite{Ando1998,Ando1998a}, and by Todorov and White \cite{White1998} who demonstrated that the chiral nature of the charge carriers in nanotubes inhibits backscattering [in all graphene structures (including graphene), chirality results from the equivalence of the A and B sub-lattices]. Ando first recognized the formal analogy between neutrino wave functions and those that describe electrons near the Fermi level in nanotubes (and in graphene). Neutrinos are massless fermions that are described by the Weyl's equation (or massless Dirac equation) \cite{Ando1998a}. The quantum number associated with chirality is the pseudospin which, like spin, can have two values. Unlike spin, the pseudospin is coupled to the momentum. In order to backscatter an electron, the scattering potential must reverse both the momentum and the pseudospin. Interactions that act equivalently on A and B atoms (like long-range potentials) conserve pseudospin and cannot backscatter charge carriers.

Ballistic conduction is only one of the favorable electronic properties of carbon nanotubes.  Others are the extremely weak electron-phonon coupling \cite{Frank1998,Hertel2000}, the excellent FET characteristics \cite{Tans1998}, and the robustness of the material itself. All of these properties indicate that nanotubes could be used for nanoelectronics. Unfortunately, incorporation of nanotubes in large-scale integrated electronic architectures proves to be so daunting that it may never be realized. Harnessing these properties requires graphitic materials that are related to carbon nanotubes, but which are more manageable.

Precisely these theoretical considerations led us in 2001 to speculate that 2D graphene could serve these purposes. We initiated experiments on epitaxially grown graphene on single crystal silicon carbide. Much of the earlier efforts focused on producing and characterizing the epitaxial graphene material. While we have achieved some success, much work remains. To fully exploit the properties of nanopatterned epitaxial graphene, one must control the graphene material, its structure, and the chemistry and morphology of defined edges.  These are the challenges for graphene-based nanoelectronics. The most important feature of 2D epitaxial graphene is that interconnected structures can in principle be patterned on the scale of an entire wafer. If, like carbon nanotubes, the carriers remain ballistic, it will lead to a fascinating world of coherent carbon-based electronics.

The discovery of the intriguing properties of deposited exfoliated graphene has recently caused overwhelming excitement in the 2D electron gas community \cite{Novoselov2005,Zhang2005,Peres2006,Gusynin2005a}. This very fascinating material clearly demonstrates the chiral nature of the charge carriers, as it manifests in several properties, of which the anomalous phase in the quantum Hall effect is the most striking. The spontaneous rippling caused by the Mermin-Wagner transition \cite{Mermin1966,Hohenberg1967,Meyer2007} and the absence of the weak anti-localization, possibly due to the gauge field at the ripples \cite{Morozov2006}, as well as the recently discovered high-field splitting of the Landau levels \cite{Zhang2006} are all very important effects that still require full explanation.

The possibility that epitaxial graphene may serve as a platform for carbon-based nanoelectronics has further greatly amplified the interest in this field, especially in the electronics community. However, epitaxial graphene and deposited exfoliated graphene are very different materials.  Epitaxial graphene is generally multi-layered whereas exfoliated graphene has only one layer. Therefore, epitaxial graphene, is a much more complex material; in fact it represents a class of materials.  It may seem that epitaxial graphene is simply ultrathin graphite, but this is emphatically not so. Experimentally, the charge carriers in epitaxial graphene are found to be chiral and the band structure is clearly related to the Dirac cone \cite{Berger2004,Rollings2006,Berger2006,Ohta2006,Sadowski2006,Wu2007}. To lowest order, epitaxial graphene appears to consist of stacked, non-interacting graphene sheets, the first of which is highly charged and the others carry much lower charge. In contrast to deposited exfoliated graphene, anomalous phase-transition-like state-changes are often observed in transport measurements of epitaxial graphene, that are probably related to weak interlayer interactions.

 These first measurements suggest that, like most layered quasi-2D conducting materials, epitaxial graphene is poised to present a host of interesting new phenomena. A snapshot of the emerging science and technology of epitaxial graphene is given here.

\section{Epitaxial graphene formation and characterization}

\begin{figure}
\centerline{\includegraphics[width=0.42\textwidth]{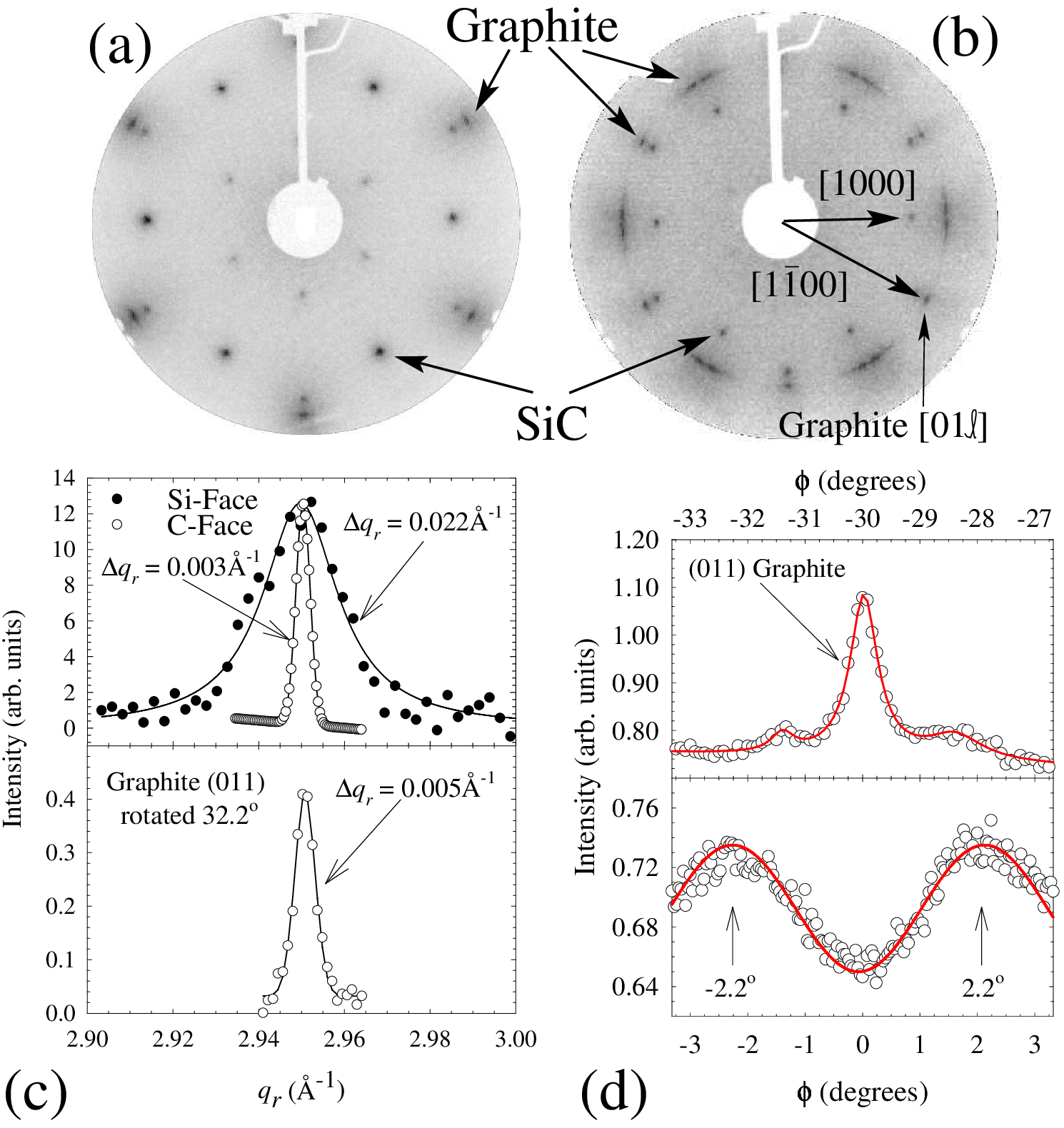}}
\caption{\label{fig:LEED} LEED and x-ray diffraction from multilayer   graphene grown on 4H-SiC(000$\overline{1}$) substrates.  (a) LEED   pattern (71 eV) for $\sim 3$ ML graphene, (b) LEED pattern (103 eV)   for $\sim 4$ ML graphene (unlabeled sets of 6-fold spots in (a) and   (b) are from a $\sqrt{3}\times\sqrt{3}R30^{\circ}$ SiC interface   reconstruction).  (c) Radial x-ray scans through (top) the   (10$\ell$) graphite rod, and (bottom) across the diffuse arcs seen   in (b).  (d) Azimuthal x-ray scans across (top) the graphite   (10$\ell$) rod and (bottom) the diffuse rods seen in (b).  }
\end{figure}

It is well known that ultrathin graphitic films grow on hexagonal silicon carbide crystals \cite{vanBommel75,Owman96,Li96a,Forbeaux00,Charrier02}. Specifically they grow on the 0001 (silicon-terminated) and $000\overline{1}$ (carbon-terminated) faces of 4H- and 6H-SiC when crystals are heated to about 1300$^\circ$C in ultra-high vacuum (UHV).
%
%
It is also possible to grow these films at more moderate vacuum conditions using ovens with controlled background gas.  The epitaxial growth is established by examining, for example, the LEED patterns after various growth times (see e.g., \rfig{fig:LEED}).  Growth on the Si face is slow and terminates after relatively short times at high temperatures. The growth on the carbon face apparently does not self-limit so that relatively thick layers ($\sim$ 4 up to 100 layers) can be achieved.

\begin{figure}
\centerline{\includegraphics[width=0.42\textwidth]{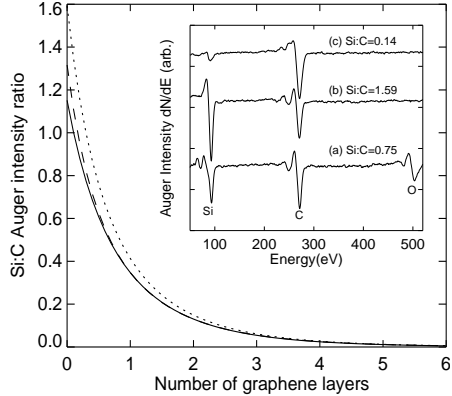}}
\caption{\label{fig:Auger} Model of Si:C Auger peak intensity ratio   versus number of graphene layers for SiC(0001) substrates. Solid   line: Model with interface layer of C adatoms at 1/3 their bilayer   density.  Dotted line: Model with interface layer of Si adatoms at   1/3 their bilayer density.  Dashed line: Model with bulk-terminated   SiC(0001). Inset shows Auger spectra obtained after (a) ex-situ   H$_2$ etching (no UHV preparation), (b) UHV anneal at   1150$^{\circ}$C (LEED $\sqrt{3}\times\sqrt{3}$ pattern), (c) UHV   anneal at 1350$^{\circ}$C (LEED $6\sqrt{3}\times6\sqrt{3}$ pattern).}
\end{figure}

For thin layers, we can estimate the graphene thickness by modeling measured Auger-electron intensities \cite{Berger2004} or photoelectron intensities \cite{Rollings2006}.  \rfig{fig:Auger} shows model results for the Si:C Auger intensity ratio for graphene grown on SiC(0001) substrates, with three different assumptions for the interface layer between bulk SiC and the graphene layers (see caption).  The Auger model, valid for both 4H and 6H polytypes, includes the relative sensitivity factors for Si and C \cite{Ertl85}, attenuation of the 3keV incident electrons and of the Auger electrons exiting from successively deeper layers \cite{Powell99,Powell91,Tanuma91}, and the electron collection angle (42$^\circ$).  Thicker multilayer graphene can be measured via conventional ellipsometry.

\begin{figure}
\centerline{\includegraphics[width=0.42\textwidth]{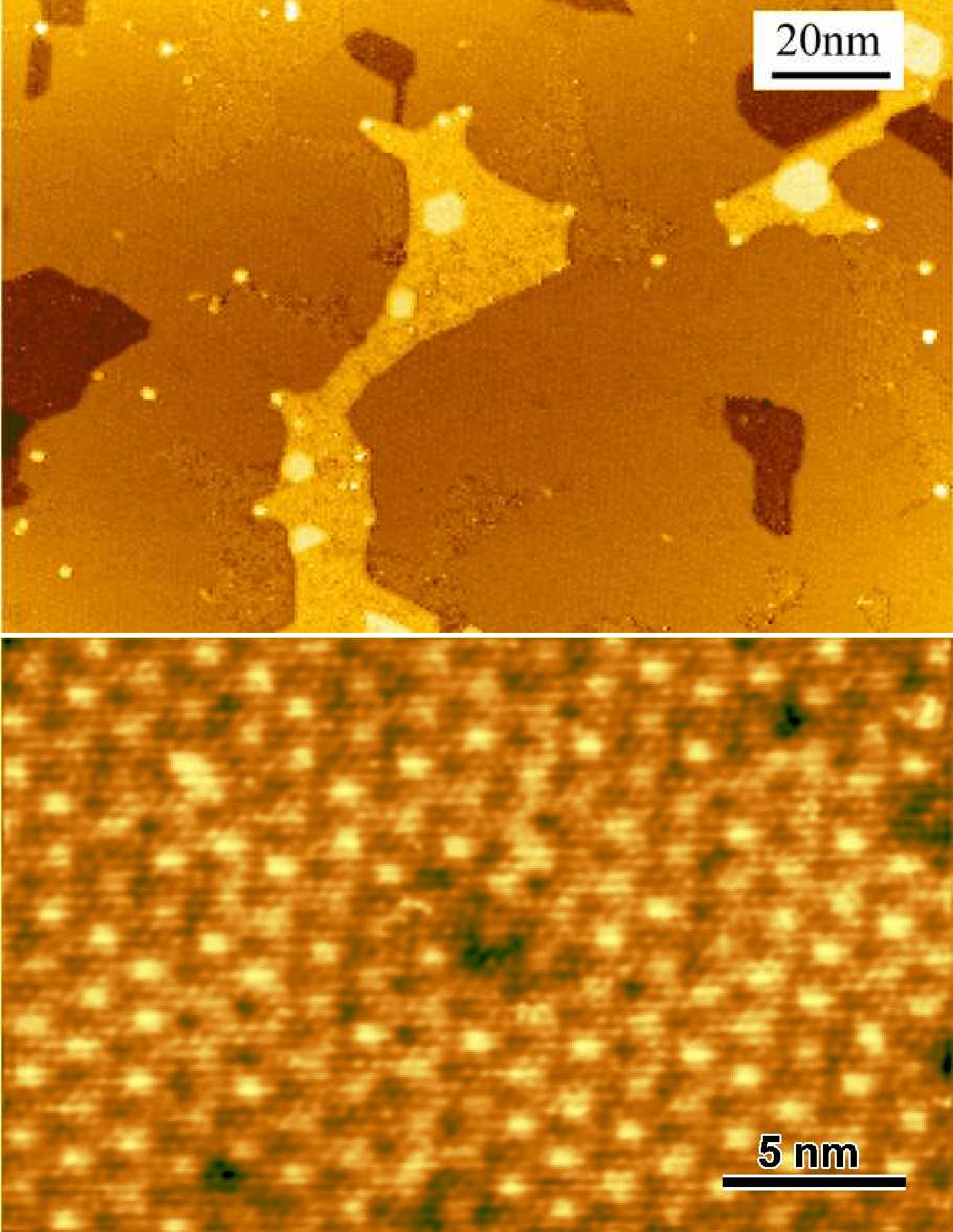}}
\caption{\label{fig:STM} STM topographs (0.8 V sample bias, 100 pA) of   nominally 1 ML epitaxial graphene on SiC(0001).  Top: Image showing   large flat regions of $6\sqrt{3}\times6\sqrt{3}$ reconstruction and   regions where the reconstruction has not fully formed.  Next-layer   islands are also seen.  Bottom: A region of   $6\sqrt{3}\times6\sqrt{3}$ reconstruction, imaged through the   overlying graphene layer.}
\end{figure}

Scanning tunneling microscopy images of monolayer graphene on the surfaces of 4H- and 6H-SiC(0001) (Si-face) show large flat regions with a characteristic hexagonal corrugation of $\sim 0.3$ \AA\  on a 1.9-nm period (\rfig{fig:STM}).  Small-scale images resolve the graphene atomic lattice throughout \cite{Berger2004,Owman96}, but with a factor 10$\times$-20$\times$ smaller amplitude.  Imaging for the monolayer is apparently dominated by interface states of an underlying reconstruction of the SiC. In conjunction with the graphene overlayer is a $6\sqrt{3}\times6\sqrt{3}$R30$^{\circ}$ reconstruction with respect to the bulk-terminated SiC surface.  The detailed reconstruction of this surface is still a matter of debate \cite{Chen05a}.  Successive graphene layers show much less influence of the interface states \cite{Berger2004}, but the 1.9-nm corrugation period ($6\times6$ with respect to the SiC bulk-terminated surface) is still visible in both STM and LEED for the thickest Si-face films we have prepared [5-6 monolayers (ML)].

To date, most transport measurements have been done on multilayer graphene grown on the carbon face [SiC(000$\overline{1}$) substrates]. This material is grown in an RF-induction furnace at pressures of $\sim 10^{-5}$ Torr.  Because the initial film-growth is very rapid, it is rare to obtain films thin enough for direct STM and LEED studies of those layers near the SiC interface.  As a consequence of charge transfer from the SiC, these layers are the most important for electrical transport.  Surface x-ray scattering has proved to be a useful tool for extracting quantitative information about the C-face-grown material.

Figure~\ref{fig:LEED} shows LEED patterns from two graphene films grown on 4H-SiC(000$\overline{1}$) substrates.  According to the Auger ratios, these were nominally (a) 3~ML graphene, and (b) 4~ML graphene. The LEED pattern in \rfig{fig:LEED}(a) shows relatively good registry to the SiC substrate (with the unit cell rotated by $30^{\circ}$, as for Si-face material), whereas the film in \rfig{fig:LEED}(b) shows some rotational disorder.  The evidence suggests that epitaxial growth does occur at the interface, but that succeeding graphene sheets do not have strong rotational order. Interestingly, the diffuse rings in \rfig{fig:LEED}(b) are clearly centered around a minimum in intensity on the SiC azimuth, indicating some preferential alignment, as discussed below.

While there is azimuthal disorder in the film, the long range vertical order of the film is much larger than is observed for Si-face grown films \cite{Hass2006}.  This is demonstrated in \rfig{fig:LEED}(c) that shows radial x-ray diffraction scans through both the graphite (10$\ell$) graphite rod ($\phi =-30.0^{\circ}$ in the [1100] SiC direction) and through the diffuse rings ($\phi = 2.2^{\circ}$ in the SiC [1000] direction).  The x-ray profiles for both the $(10\ell=1.5)$ and diffuse rods on the C-face graphene are nearly 10 times narrower than those for Si-face films.  The profile widths are inversely related to the size of order graphene domains; $L=2\pi/\Delta q_r$. For Si-face films the order graphene regions are $\sim 290\textrm{\AA}$ while for the C-face films the domains are $\sim 2100\textrm{\AA}$.  The domain size estimated this way is most likely a lower limit on the actual size of a graphene sheet.  A continuous graphene sheet (typically 3000\AA\ terrace width) folded over a SiC step would break the scattered x-ray coherence from the two regions, but may have a much smaller influence on the electronic structure. Note that even the diffuse rings have domain sizes of $\sim 1200\textrm{\AA}$.

In fact, the rotationally disordered graphene has a structure.  \rfig{fig:LEED}(d) shows x-ray azimuthal scans through both the graphite $(10\ell)$ and diffuse graphite rods.  The diffuse rings are in fact peaked at $\pm 2.2^{\circ}$ relative to the SiC azimuth.  This angle is not arbitrary.  It corresponds to a structure were two vertically stacked graphene sheets are commensurate if rotated with respect to one another by $\cos 11/13 =32.204^{\circ}$ \cite{Kolmogorov05}. Both $30^{\circ}$ and $\pm 2.204^{\circ}$ rotated graphene are also nearly commensurate with the SiC $6\sqrt{3}\times6\sqrt{3}$ R30 seen in Si-face grown graphene [see \rfig{fig:LEED}(a)] It therefore seems that during graphitization large graphene sheets are free to rotate with respect to each other and lock in, on average, to these preferred orientations on the SiC C-face.

In addition to the difference in long range and orientational order of films grown on the two polar faces of SiC, the vertical roughness of the multilayer graphene is very different.  X-ray diffraction reveals that the rms roughness of the C-face multilayer films is less than $0.05$~\AA\ over the 2~$\mu$m coherence length of the beam \cite{Hass2007}.  On the Si-face the roughness is much larger ($\sim 0.2$\AA\ \cite{Hass07a}), presumably as a consequence of the $6\times 6$ corrugation (see \rfig{fig:STM}).

Finally, x-ray reflectivity experiments show two other important features of multilayer graphene grown on the C-face of SiC. First, the first layer of graphene sits 1.62\AA\ above the last SiC layer \cite{Hass2007,Varchon07}.  This bond length is nearly equal to the bond length of diamond ($1.54\textrm{\AA}$) and suggest that the substrate bond to the first graphene layer is much stronger than a van der Waals interaction.  In fact {\it ab intio} calculations find and very similar bond distance \cite{Varchon07}.  These calculations show that the first graphene layer is in fact insulating. Only the formation of the second graphene layer gives rise to an electron dispersion curve showing a Dirac cone.  Thus the first graphene layer can be interpreted as a ``buffer'' layer between the substrate and an isolated layer with the electronic properties of an isolated graphene sheet.

The second important result from the x-ray reflectivity is that the graphene interlayer spacing is significantly larger than bulk graphite \cite{Hass2007}.  The measured value is $3.368$\AA\ which is between the value of bulk graphite and turbostratic graphite.  This larger spacing suggest a significant density of stacking faults.  This is not too surprising given the rotational disorder in the C-face films.  For a random stacking fault model the layer spacing can be used to estimate the stacking fault density to be one every other layer \cite{Hass2007,Franklin51}.  This type of density suggest that the \emph{AB} stacking order, that would destroy the graphene electronic character, is nearly lost in these films and may significantly impact the transport properties of these films.

\section{Landau level spectroscopy of epitaxial graphene}

\begin{figure}
\centerline{\includegraphics[width=0.8\textwidth]{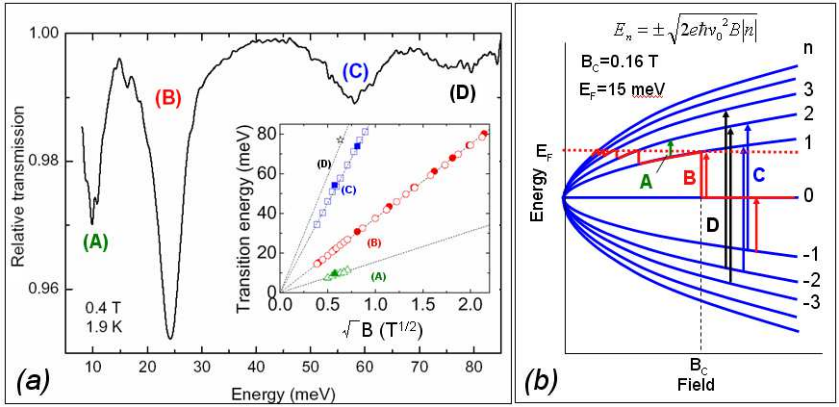}}
\caption{\label{IRSpec} Infrared transmission spectroscopy of epitaxial graphene with about 10 layers revealing Landau level structure. (a) Infrared transmission spectrum at $B=0.4$ T and $T=1.9$ K, showing a series of absorption peaks. (Inset) The absorption maxima positions as a function of field showing the $\sqrt{B}$ dependence that is characteristic for a chiral ``massless" Dirac particle. (b) Schematic diagram of the Landau levels $E_n(B)$ in which the only parameter is $v_0$ that is found to be 10$^8$ cm/s. The arrows indicate the observed transitions. $E_F$ is determined from the lowest field for which the $n=0$ to $n=1$ transition is observed. }
\end{figure}

Dirac particle properties of the charge carries in epitaxial graphene multilayers have been beautifully demonstrated in Landau Level spectroscopy by Sadowski \textit{et al.} \cite{Sadowski2006}. (See Sadowski {\it et. al.} in this issue for a summary and update). We summarize some of the results here. In these measurements, an epitaxial graphene sample is illuminated by infrared light in a magnetic field at low temperatures. The absorption is measured as a function of photon energy at various magnetic field strengths. An example of such a spectrum is shown in \rfig{IRSpec}. The various absorption lines are identified as transitions between various Landau levels. The transitions energy are found to accurately follow $E_n=v_0\sqrt{2ne\hbar B}$. The exact $\sqrt{B}$ dependence is the hallmark of a "massless" Dirac particle (more precisely, of a linear density of states); massive particles have a linear B dependence. Moreover, a gap at the tip of the Dirac cone also distorts the $\sqrt{B}$ behavior. The Fermi velocity is determined from the dispersion of the transitions with magnetic field to be $v_0=1.03\times10^8$ cm/s, which is close to its value for exfoliated graphene. The $n=0$ to $n=1$ transition is observed only for $B\geq0.16$ T, which indicates that the $n=1$ level is just depopulated at that field. Hence, -15 meV $<E_F<$ 15 meV and $n\approx1.5\times10^{10}$ /cm$^2$ and the Fermi wavelength is $\approx300$ nm. It is further found that the intensity of the signal scales with the thickness of the film. These experiments demonstrate that epitaxial graphene consists of stacked graphene layers, whose electronic band structure is characterized by a Dirac cone with chiral charge carriers. Remarkably, there is no evidence for a gap nor for a deviation of the linear density of states: undistorted Dirac cone properties are directly observed as close as 20 meV to the Dirac point in the $n=0-n=1$ transitions.

Epitaxial graphene is clearly not graphite, which has a different spectrum and an entirely different electronic structure (see Sadowski \textit{et al.} in this issue). This difference reflects that epitaxial graphene does not have the Bernal stacking that would lift the pseudospin degeneracy \cite{Hass2007}. Hence epitaxial graphene is a form of multilayered graphene that is structurally and electronically distinct from graphite. These experiments probe the low charge density bulk of the epitaxial graphene layer. Below we discuss the highly charged interface layer.

\section{Patterning epitaxial graphene }

\begin{figure}
\centerline{\includegraphics[width=0.4\textwidth,angle=-90]{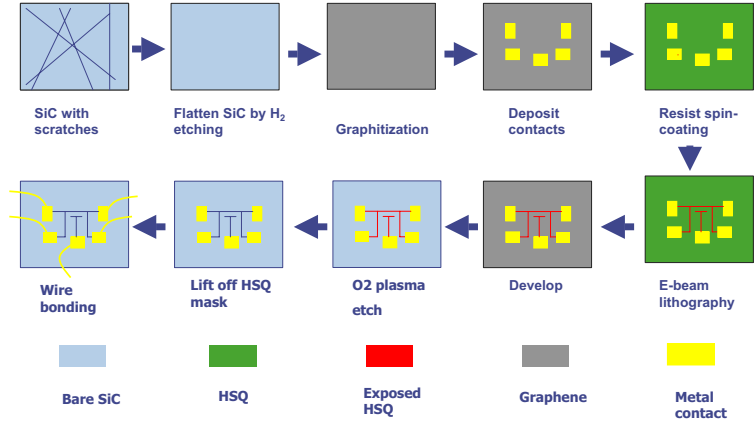}}
\caption{\label{patterning} Patterning epitaxial graphene}
\end{figure}

Epitaxial graphene samples are patterned using a variety of microelectronics patterning methods. Features down to several tens of nanometers are produced by standard e-beam lithography methods. The method is outlined in \rfig{patterning}.

\section{Transport in 2D epitaxial graphene}

\begin{figure}
\centerline{\includegraphics[width=0.6\textwidth]{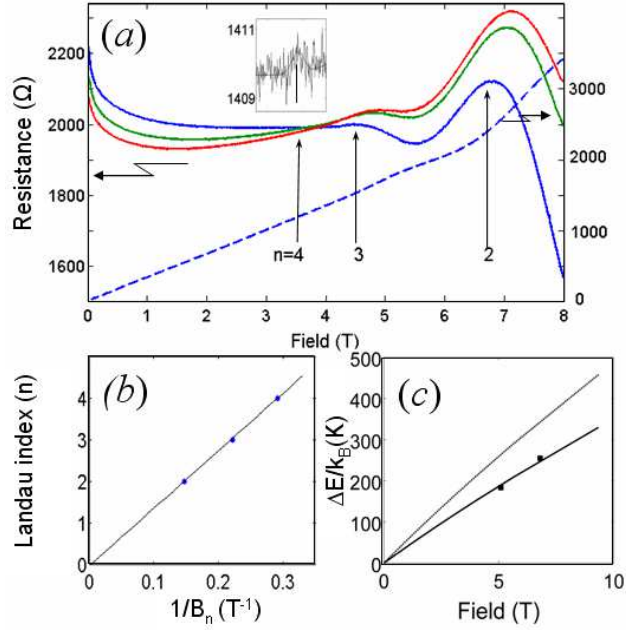}}
\caption{\label{2D-EG-Si} 2D transport measured in a 400 $\mu$m by 600 $\mu$m Hall bar on 3 layer epitaxial graphene on the Si face. Mobility $\mu=1200$ cm$^2$/V$\cdot$s, coherence length $l_\phi=300$ nm. (a) Magnetoresistance at $T=$0.3, 2 and 4 K showing well developed SdH peaks, indicated with their Landau indices $n$; the Hall resistance at 0.3 K (dashed line), shows a weak feature at the expected Hall plateau position. The amplitude of the weak localization peak at $B=0$ corresponds to $1G_0$. (b) Landau plot; the linear extrapolation passes through the origin demonstrating the anomalous Berry's phase characteristic of graphene. (c) The Lifshitz-Kosevich analysis of the $n=2$ and $n=3$ peaks which correspond to graphene with a Fermi velocity $v_F=7.2\times10^5$ cm/s.}
\end{figure}

\begin{figure}
\centerline{\includegraphics[width=0.6\textwidth]{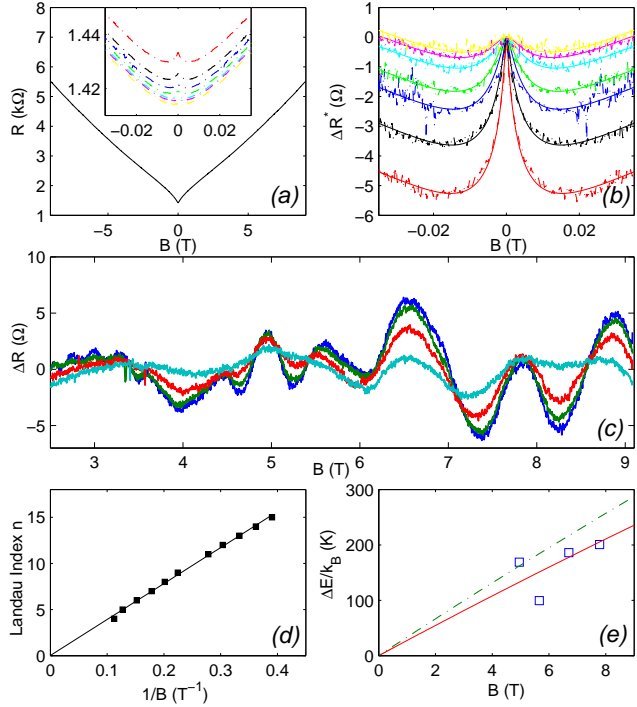}}
\caption{\label{2D-EG-C} 2D transport in a 100 $\mu$m $\times$ 1000 $\mu$m Hall bar on a $\sim$10 layer eptitaxial graphene film on the C face. a) Resistance as a function of the magnetic field. Inset, dash-dot lines, low field MR at various temperatures (1.4, 4.2, 7, 10, 15, 20, 30, 50 K). (b) Low field MR after subtracting 50 K data as a background. dash-dot lines, experimental data, which show suppressed weak localization peak around zero. The positive MR above 0.02 T reveal the weak anti-localization effect. Solid lines, fits to the theory by McCann \textit{et al.}. (c) High field MR after subtracting a parabolic background at several temperatures(4, 7, 15, 30 K). Well defined SdH oscillations can be seen down to 2.5 T. (d) Landau plot for SdH oscillations, which intercept $y$ axis at zero. (e) Landau level spacing obtained by Lifshitz-Kosevich analysis. Squares: experiment. Solid line, theoretical prediction for $\Delta E$ assuming $v_F=0.82\times10^8$ cm/s, dash-dot line: $v_F=10^8$ cm/s.}
\end{figure}

The first published transport measurements on epitaxial graphene were made on a Hall bar patterned on a graphene film with about 3 layers on the silicon face of 4H-SiC \cite{Berger2004}. The mobility of the sample was relatively low (1100 cm$^2$/V$\cdot$s) nevertheless the Shubnikov-de Haas oscillations are clearly distinguished (see \rfig{2D-EG-Si}) \cite{Note01}. Resistance maxima in graphene are expected at fields $B_n$ when the Fermi energy intercept the Landau levels, i.e. for $E_F=v_0\sqrt{2ne\hbar B_n}$, where $v_0\approx10^8$ cm/s is the Fermi velocity, hence $B_n=(E_F/v_0)^2/2ne\hbar=B1/n$. For normal electrons maxima are found when $E_F=(n+1/2)eB_n\hbar/m$, hence $B_n= E_F m/(n+1/2)\hbar e$. Therefore the Landau plot (a plot of $n$ versus $1/B_n$) of a Dirac particle intercepts the origin whereas the Landau plot of a normal electron intercepts the y axis at $n=1/2$. The intercept should occur at 0 when the Berry's phase is anomalous. This shows that the Landau plot provides a ready method to identify a Dirac particle when the quantum Hall measurements are not feasible. The Landau plot (\rfig{2D-EG-C}) for data on a sample similar to the one of \cite{Berger2004} passes through the origin indicating that the Berry's phase is anomalous. The Hall coefficient at 0.3 K is found to be 330 $\Omega$/T corresponding to a charge density of $2\times10^{12}$ electrons/cm$^2$. (Note that for a Dirac particle it should be $6500/B_1=450$ $\Omega$/T.) From $v_0=10^8$ cm/s we further find that $E_F\approx1680$ K. The large charge density is caused by the built-in electric field at the SiC-graphene interface, which dopes the interfacial graphene layer. This layer carries most of the current (and causes the SdH oscillations). The charge density of the top layers is more than 2 order of magnitude smaller (see above) and they are expected to be much more resistive. The temperature dependence of the SdH peak amplitudes is determined by the Landau level spacing $E_{n+1}(B)-E_n(B)$ and given by the Lifshitz-Kosevich equation: $A_n(T)\sim u/\sinh(u)$ where $u=2\pi k_B^2T/\Delta E(B)$ \cite{Lifshitz1956}. From this fit we find that at $B=7$ T, $(E_3(B)-E_2(B))/k_B=250$ K (compared with 340 K predicted for graphene at this carrier density)  and that the Dirac point is about 1290 K below $E_F$.

This sample shows ample evidences that the carriers in the high-charge-density layer, like those in the low-density layers, are Dirac electrons.  However the quantum Hall effect is not observed. Instead, only weak undulations are seen in the Hall resistance. It was assumed that higher mobility samples would enhance the QHE and subsequent work progressed in that direction. Note also the intense weak localization peak near $B=0$ indicative of significant point-defect scattering. Due to the high current density, the interface graphene layer dominates the transport, although the other layers are expected to contribute, and more so in 2D structures than in quasi 1D structures (see below).

Graphene grown on the Si face typically has low electron mobilities. The very thin films are relatively unprotected from even slight residual oxidizing gases that damage the graphene \cite{Hass2006}. Work is still progressing to improve Si face graphene films.

On the other hand, graphene grown on the C face has much higher mobilities \cite{Berger2006}. The films are also considerably thicker so that the high-density layer at the interface is more protected \cite{Hass2007}. \rfig{2D-EG-C} shows the MR measurements of a Hall bar (100 $\mu$m $\times$1000 $\mu$m) at several temperatures \cite{Wu2007}. The SdH oscillations are barely discernable, which is generally the case for our high mobility 2D samples. The reason for this is not likely due to sample inhomogeneity. The Landau plot of the oscillations reveals the anomalous Berry's phase, characteristic of Dirac electrons. Furthermore the charge density is $3.8\times10^{12}$ electrons/cm$^2$. The charge density from the Hall effect is $4.6\times10^{12}$ electrons/cm$^2$. The Lishitz-Kosevich analysis of the peak heights agrees with the expected Landau level spacing for a Dirac particle.

A striking feature of this sample is that the weak localization peak is very weak, $\sim0.07G_0$ (compared with the  sample in \rfig{2D-EG-Si}) which indicates that point defect density in this sample is low and these defects are  possibly localized entirely at the patterned edges of the Hall bar. On the other hand, a marked temperature dependent  depression of the conductance at low fields is observed. This feature suggests weak anti-localization that is expected  when Dirac electrons are scattered by long-range potentials \cite{Ando1998,Ando1998a}. These could be due to the  localized counterions in the SiC substrate. In fact the amplitude, field and temperature dependence of this feature  match predictions of the weak anti-localization very well \cite{McCann2006}.

Another typical feature is the large positive magnetoresistance and a kink in the Hall resistance at low fields. These  features (as well as the small discrepancy in the charge density) could be due to the other layers of density $n\lesssim10^{10}$ /cm$^2$ \cite{Sadowski2006}, although no SdH features can be attributed to them. It should be noted that the critical field $B_c$ for which extreme quantum limit is  reached (where $E_F$ coincides with the $n=0$ Landau level, i.e. at about 30 meV above the Dirac point) is also very  low: $B_c\leq160$ mT (see \rfig{IRSpec}).

The Hall resistance is featureless (except for extremely weak ripples) and shows no evidence for quantum Hall plateaus,  as is for a typical high mobility 2D samples.

\begin{figure}
\centerline{\includegraphics[width=0.6\textwidth]{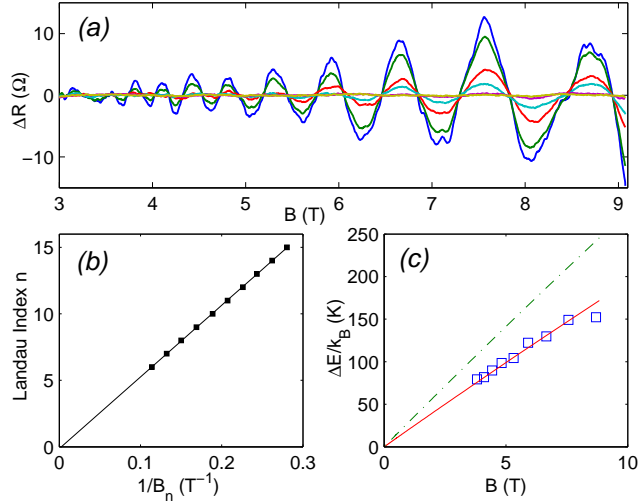}}
\caption{\label{q1D-EG} Intermediate width Hall bar:  1 $\mu$m $\times$ 5 $\mu$m. The zero field resistance is 502 $\Omega$. (a) High field MR after subtracting a smooth background at several temperatures(4, 10, 20, 30, 50, 70 K). (b) Landau plot. $B_1=53$T, intercept 0.13$\pm$0.02 (c) Square: Landau level spacing $\Delta E$ obtained by fitting the temperature dependence of SdH amplitudes to LK equation. Solid line, theoretical prediction for $\Delta E$ assuming $v_F=0.7\times10^8$ cm/s, dash-dot line: $v_F=10^8$ cm/s}
\end{figure}

The transport properties of a narrower ribbon are shown in \rfig{q1D-EG}. It is at once clear that the SdH oscillations are much more pronounced. The Landau plot corresponds quite well with the expectations for a Dirac particle with a velocity $0.7\times10^8$ cm/s. This ribbon shows evidence for weak anti-localization. A more pronounced weak localization peak compared with \rfig{2D-EG-C} is observed. However the Hall resistance, which is quite similar to that in \rfig{2D-EG-Si}, shows no evidence for the QHE.

\section{Transport in quasi-1D epitaxial graphene}

\begin{figure}
\centerline{\includegraphics[width=0.5\textwidth]{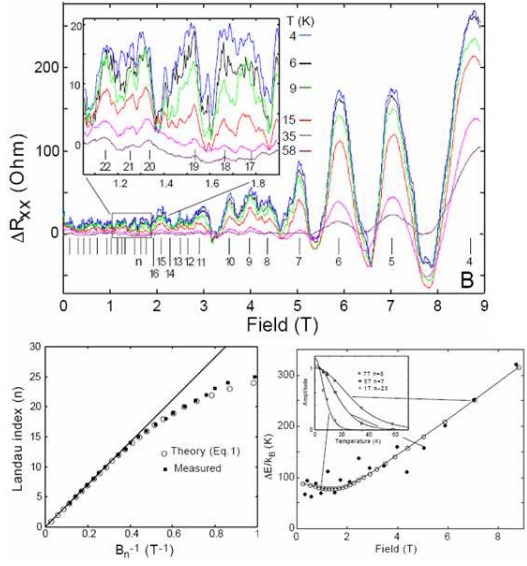}}
\caption{\label{1D-EG} Narrow Hall bar 500 nm $\times$ 6 $\mu$m. The zero field resistance is 1125 $\Omega$. (a) Magnetoresistance oscillations for temperatures ranging from 4-58 K after subtraction of a smooth background. (b) Landau plot of the magnetoresistance peaks. The deviation for large from linearity is due to quantum confinement. (c) The energy gap between the Fermi level and the lowest unoccupied Landau level is found from the Lifshitz-Kosevich analysis (inset) of the peaks and increases linearly with field for large fields and saturates for low fields. The saturation confirms quantum confinement.}
\end{figure}

Quantum confinement effects manifest in narrow ribbons. As for 2D Hall bars, this interface graphene layer is charged with about $4\times10^{12}$ electrons/cm$^2$ which corresponds to  a Fermi wavelength of about 20 nm. Since the Fermi wavelength of the low-density layers is about 400 nm, consequently for ribbons that are narrower than 500 nm, these layers contribute little to the transport. For very narrow ribbons  ($\leq100$ nm) with rough edges, the low-density layers are expected to be insulating, since there are no propagating modes (channels).

Figure~\ref{1D-EG} shows the Hall resistance and the magnetoresistance of a narrow ribbon (see Ref. \cite{Berger2006} for details). The Landau levels for a graphene ribbon are approximately given by
\begin{equation}
E_n(B,W)\approx[E_n(W)^4+E_n(B)^4]^{1/4}
\end{equation}
where $E_B(n)=\sqrt{2neB v_0^2\hbar}$ and $E_W(n)=n\pi\hbar v_0/W$ \cite{Peres2006a}. Confinement effects become apparent for low fields, approximately when the cyclotron diameter becomes greater than the ribbon width. Confinement will then cause deviation from the linearity in the Landau plot as seen in \rfig{1D-EG}. The Lifshitz-Kosevich analysis confirms the confinement. For high magnetic fields the energy separation between the Landau levels increases with increasing field as expected, while for low field the energy separation saturates and is determined by the quantum confinement. Note that this analysis does not require a determination of the locations of the magnetoresistance peaks (Ref. \cite{Berger2006}).

The mobilities of the graphene ribbons appears to increase with decreasing width, \rfig{mobility}. This effect may be related to the reduced back-scattering with decreasing number of conducting channels. On the other hand, back-scattering at the ribbon edges should become relatively more important with decreasing width. The amplitudes of the SdH oscillations are much more pronounced for narrow ribbons than for high mobility 2D Hall bars.

\begin{figure}
\centerline{\includegraphics[width=0.5\textwidth]{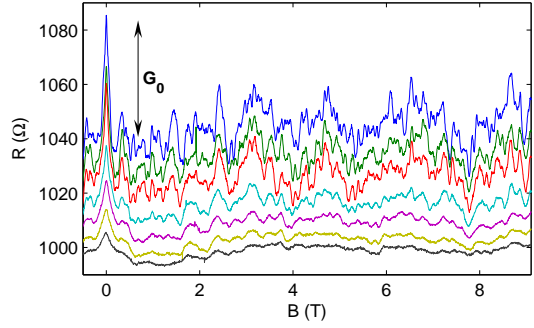}}
\caption{\label{fluctuation} Magnetoresistance of a 0.2 $\mu$m $\times$ 1 $\mu$m ribbon. The experiment were done at 4, 8, 12, 30, 45, 60, 90 K, from top to bottom. The resistance has been shifted for clarity, except for 4 K. The amplitude of the weak localization peak at zero field is about $1G_0$. }
\end{figure}

A relatively large fraction of the high-mobility narrower Hall bar samples do not exhibit SdH oscillations at all, as
seen in \rfig{fluctuation}. Occasionally rather complex magnetoresistance structures that in many cases appear not to be random but exhibit features that are approximately linear in field (like in the Aharonov-Bohm effect). Several of these
systems are found to be coherent and ballistic. In one case the resistance of a $0.5\times5$ $\mu$m Hall bar abruptly and reversibly drops by an order of magnitude at $T=200$ K to below 10 $\Omega$/sq. It appears that scattering at the edges is specular without any back-scattering. The effects point to a correlated electronic system (Levy, Berger, de Heer \textit{et al.}, to be published).

\section{Structure dependent properties and the absence of the quantum Hall effect}

\begin{figure}
\centerline{\includegraphics[width=0.7\textwidth]{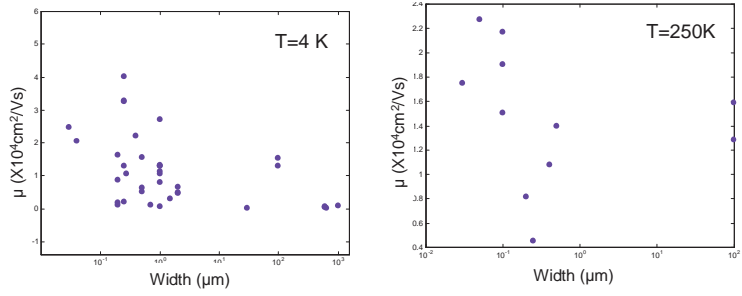}}
\caption{\label{mobility} The width dependence of mobility.}
\end{figure}

A key focus of epitaxial graphene research is to develop a new graphene-based electronics material with shape tunable properties. The intrinsic width dependent bandgap of graphene ribbons has been born out experimentally in back-gated deposited exfoliated graphene ribbons \cite{Han2007}. We have not yet demonstrated the effect in epitaxial graphene, primarily due to problems in gating the material, which we hope to solve soon.

Currently we have reasonable statistics that appear to suggest that the mobilities of the ribbons actually increase with decreasing ribbon width (\rfig{mobility}). This intriguing property could be due to the fact that the system becomes more one-dimensional with decreasing width and thereby that backscattering is inhibited. On the other hand, the decreasing width also implies that the edges (which are presumed to be rough) become more important and enhance the scattering. Apparently that effect is not dominant.

It is remarkable that the SdH oscillations are extremely weak except for very low mobility samples, that are known to be quite defective (as in \rfig{2D-EG-Si}). In fact the SdH oscillations are almost imperceptible in the 2D sample (the amplitudes are only 0.001 of the mean resistance) even though they are well resolved up to the 15th Landau level. The weak localization peak is weak ($\sim 0.07 G_0$) and evidence is seen for weak anti-localization. In contrast, the oscillation of the 2nd Landau level in the low mobility sample 2D is large (0.3 of the mean resistance); this sample may exhibit the quantum Hall effect at high fields. Furthermore, the weak localization peak is intense ($\sim 1 G_0$) In the intermediate regime, the 1 $\mu$m width ribbon exhibits well resolved SdH peaks (0.016 of the mean resistance) while the weak localization peak is $0.52 G_0$, weak anti-localization is also present.

Narrow ribbons exhibit more intense weak-localization peaks, well-resolved SdH oscillations, quantum confinement peaks, and high mobilities but no evidence for the quantum Hall effect. It may be assumed that the QHE in the high-density layer is shorted out by the low density layers, however this is not bourn out in simulations. For example, it is not possible to "convert" the oscillations of \rfig{1D-EG} to those of \rfig{2D-EG-C} by adding the conductivity of many graphene layers to the former. Note that the relative SdH oscillation amplitudes in \rfig{2D-EG-C} are 16 times smaller than in \rfig{q1D-EG}, while they are more than 20 times smaller in \rfig{2D-EG-Si}, while the square resistances of all three are within a factor of 3 from each other.

The fact that the most intense SdH peaks in 2D samples are seen in the most defective samples, leads us to conclude that defects, specifically in the "bulk" of the sample (i.e. away from the edges) are required for large amplitude SdH peaks, and hence for the QHE.

This point of view is strengthened by the fact that a coulomb (electrostatic) potential cannot trap Dirac particles \cite{Klein1929,Martino2007}. Hence, if scattering away from the edges is primarily from (long-range) coulomb potentials due to counter ions in the SiC substrate, then these potentials cannot trap the carriers. It is well known that localized states in the bulk are required for the QHE so that the absence of such states would inhibit the QHE \cite{Yoshioka2002,Ilani2004}! It would be of course very important that this conclusion is verified since it so dramatically departs from observations in deposited exfoliated graphene samples, which further underscores fundamental differences in these materials.

\bibliographystyle{elsart-num}
\bibliography{EG}





\end{document}